\documentclass[sigconf]{acmart}
\AtBeginDocument{%
  }

\copyrightyear{2025}
\acmYear{2025}
\setcopyright{acmlicensed}\acmConference[ICTIR '25]{Proceedings of the 2025 International ACM SIGIR Conference on Innovative Concepts and Theories in Information Retrieval (ICTIR)}{July 18, 2025}{Padua, Italy}
\acmBooktitle{Proceedings of the 2025 International ACM SIGIR Conference on Innovative Concepts and Theories in Information Retrieval (ICTIR) (ICTIR '25), July 18, 2025, Padua, Italy}
\acmDOI{10.1145/3731120.3744602}
\acmISBN{979-8-4007-1861-8/2025/07}

\usepackage{xcolor} 
\usepackage{multirow}

\usepackage{balance}

\begin{document}

\title{Impact of Shallow vs. Deep Relevance Judgments on BERT-based Reranking Models}

\author{Gabriel Iturra-Bocaz}
\email{gabriel.e.iturrabocaz@uis.no}
\orcid{0009-0001-9635-0683}
\affiliation{%
  \institution{University of Stavanger}
  \city{Stavanger}
  \country{Norway}
}

\author{Danny Vo}
\email{d.vo@stud.uis.no}
\orcid{0009-0006-1407-9743}
\affiliation{%
  \institution{University of Stavanger}
  \city{Stavanger}
  \country{Norway}
}

\author{Petra Galu\v{s}\v{c}\'{a}kov\'{a}}
\email{petra.galuscakova@uis.no}
\orcid{0000-0001-6328-7131}

\affiliation{%
  \institution{University of Stavanger}
  \city{Stavanger}
  \country{Norway}
}

\renewcommand{\shortauthors}{Gabriel Iturra-Bocaz, Danny Vo, and Petra Galuščáková}
\begin{abstract}
      This paper investigates the impact of shallow versus deep relevance judgments on the performance of BERT-based reranking models in neural Information Retrieval. Shallow-judged datasets, characterized by numerous queries each with few relevance judgments, and deep-judged datasets, involving fewer queries with extensive relevance judgments, are compared. The research assesses how these datasets affect the performance of BERT-based reranking models trained on them. The experiments are run on the MS MARCO and LongEval collections. Results indicate that shallow-judged datasets generally enhance generalization and effectiveness of reranking models due to a broader range of available contexts. The disadvantage of the deep-judged datasets might be mitigated by a larger number of negative training examples. 
\end{abstract}



\maketitle

\section{Introduction}

In neural Information Retrieval (IR), the method of collecting training data has a significant impact on the performance of retrieval models trained on this data. The quality and structure of the training data influence how well the models learn to distinguish relevant from non-relevant documents during the inference. Data collection approaches in IR (either for training or evaluation) which use documents manually judged for their relevance to the queries, can be generally categorized into two primary sets: \textbf{deep} and \textbf{shallow}. Deep-judged datasets contain fewer queries, each with numerous relevance judgments. These datasets, often created through pooling in IR evaluations, typically consist of tens of queries, each associated with hundreds of relevance judgments. In contrast, shallow-judged datasets use many queries, each with few relevance judgments, offering broader coverage but less depth per query. Shallow-judged datasets are often derived from real-world queries in production systems and are often used for both training and evaluation. 
For simplicity, we refer to deep-judged datasets as deep datasets and shallow-judged datasets as shallow datasets throughout this paper. The same abbreviated terminology applies to judgments (deep judgments and shallow judgments) collected on deep and shallow datasets.

Shallow and deep datasets were introduced by \citet{Yilmaz:2009:sigir}, who explored the trade-off between the number of queries and the number of judgments per query for training learning-to-rank systems. In this paper, we re-examine the effectiveness of shallow versus deep judgments, for training a BERT-based reranker \cite{Devlin:2019:arXiv:BERT}. Rather than relying solely on the statistical pattern learning used in traditional learning-to-rank, we use BERT-based rerankers which capture semantic relationships in query-document pairs  during the training phase. We additionally study the role of negative sampling. 


The central research question of this paper is: \textit{what is the impact of training rerankers on shallow datasets versus deep datasets?} While shallow judgments are expected to have an advantage due to the larger number of diverse contexts, we aim to determine if, and under what circumstances, deep judgments can outperform shallow ones. We also focus on the effect of the negative relevance judgments, as they are typically much easier to acquire than the true positive judgments by using automatized methods. Since creating relevance judgments is time-consuming and resource-intensive, determining whether manual efforts in building evaluation datasets can be effectively reused for training could have a significant impact on the IR community. We conduct our comparison using two MS MARCO collections and a LongEval collection, all of which contain shallow and deep judgments\footnote{Our code is available at https://github.com/iai-group/ictir2025-relevance}.

In the next section, we review related work, focusing on the impact of training data on reranking model performance. Section~\ref{sec:datasets} describes the collections used in the experiments, while Section~\ref{sec:experiments} outlines the experiments and analyzes the results. We discuss the results in Section~\ref{sec:discussion}.
Finally, Section~\ref{sec:conclusion} concludes the paper.

\section{Related Work}
\label{sec:related}
The introduction of BERT by \citet{Devlin:2019:arXiv:BERT} marked a significant milestone in IR, allowing models to capture complex semantic relationships within text \citep{Yates:2021:WSDM}.
This advancement paved the way for BERT's application in reranking, where it refines results obtained from traditional retrieval methods like BM25. The effectiveness of BERT as a reranker was first demonstrated by \citet{Nogueira:2019:ArXiv}. They introduced two reranking setups, MonoBERT and DuoBERT, which transformed the ranking task into pointwise and pairwise classifications, respectively. In our experiments, we adopt the MonoBERT setup.

Nogueira et al.~\cite{Nogueira:2019:ArXiv, Nogueira:2019:ArXiv:Bert, Nogueira:2020:ACL} further explored the dependence of the MonoBERT reranker on the number of training instances (query / document pairs). They found that approximately 10,000 training instances are required to outperform the baseline BM25 model on the MS MARCO collection. Like us, they fine-tuned the vanilla BERT model. Other experiments, such as those by \citet{Zhang:2020:SustaiNLP}, show that BERT pre-trained on the full MS MARCO collection can outperform BM25, even on out-of-domain data. However, fine-tuning this pre-trained BERT on too small in-domain dataset can degrade performance. 
\citet{Althammer:2023:SIGIR-AP} explored active learning strategies for the effective selection of training data for neural models. Although they observed a significant effect of training data selection on model performance, they did not find any active learning strategy that outperformed random selection of training instances. Additionally, the selection of negative samples also proved to have a significant effect on retrieval quality~\citep{Cai:2022:CIKM,Qu:2021:NAACL,Xiong:2021:ICLR}. 
Our work is mostly related to the work of \citet{Yilmaz:2009:sigir}, who studies the differences between \textbf{deep} and \textbf{shallow} relevance judgments in learning-to-rank systems on which they show that the shallow judgments are more cost effective than deep judgments. Shallow and deep judgments were also studied when used in the evaluation. \citet{Zobel:1998:SIGIR} examined the optimal pool depth and claimed that the pool depth of 100 documents might be enough for acquiring reliable results. 
\citet{Lu:2017:SIGIR} explored  to which extent can deep judgment-based methods be  effective for evaluating shallow-based judgments. They propose novel methods for such evaluation, based on a prediction of a gain of unjudged documents. 

\section{Datasets}
\label{sec:datasets}

We utilize both shallow and deep judgments from the MS MARCO and LongEval collections. The use of the terms deep and shallow here refers to how training data is gathered. Specifically, to the trade-off between query coverage and labeling approach between the number of queries vs the number of relevance judgment per query. In this context, a deep dataset is constructed by collecting many relevance judgments for each individual query, resulting in thorough labeling for a small set of queries. This approach provides insight into the relevance judgment landscape for each query but limits the diversity of queries represented in a training set. On other hand, a shallow dataset is built by collecting only a few relevance judgments per query, but considering a much broader set of queries, which increases the amount of query coverage in the dataset, exposing the BERT-based reranker model to a larger variety of information needs, but with less detailed information for each one.

\subsection{MS MARCO}

\begin{table*}
\begin{tabular}{ll|cccc|cccc}
\hline
& &\multicolumn{4}{c}{\textbf{MS MARCO V1}} & \multicolumn{4}{c}{\textbf{MS MARCO V2}}\\
\textbf{Type} & \textbf{Collection} & \textbf{Query} & \textbf{Qrels} & \textbf{PQrels} & \textbf{Docs} & \textbf{Query} & \textbf{Qrels} & \textbf{PQrels} & \textbf{Docs}\\
\hline
\multirow{4}{*}{\textbf{Deep}} & TREC 2019 DL &43&9,260&4,102&8,841,823&-&-&-&-\\
& TREC 2020 DL &54&11,386&3,606&8,841,823&-&-&-&-\\
& TREC 2021 DL &-&-&-&-&53&10,828&6,490&138,364,198 \\
& TREC 2022 DL &-&-&-&-&76&386,416&99,957&138,364,198\\
\hline
\multirow{2}{*}{\textbf{Shallow}} & Train & 502,939 & - & 532,761 & 8,841,823 & 277,144 & - & 284,000 & 138,364,198\\
& Test & 4,281 & - & 4,655 & 8,841,823 & 3,903 & - & 4,009 & 138,364,198\\ 
\hline
\end{tabular}
\caption {Statistics of the MS MARCO collections used in the experiments. Number of queries (Query), all relevance judgments (Qrels), positive relevance judgments (PQrels) and documents (Docs) are displayed. Dev collections are used as the Shallow Test collections.}
\label{tab:msmarco:stats}
\end{table*}

MS MARCO is a collection originally focused on machine reading comprehension which became a standard collection for training and evaluating neural IR.
To create the \textbf{shallow} datasets, we sample queries and passages from the Train sets of the MS MARCO V1 and V2 Passage Ranking collections. The statistics of the collections are displayed in Table~\ref{tab:msmarco:stats}. The collection is provided in the form of triples of the query, positive and negative relevance judgment. While positive judgments were collected from the human annotators, negative judgments were created automatically and we do not use them in the presented experiments. In both V1 and V2 Train sets, there is average one relevant passage per query. 
The \textbf{deep} datasets are based on the TREC Deep Learning (DL) Track datasets. These consist of the same set of passages as the shallow collection, selected queries, and a pool of manually evaluated passages for each query, returned by the models participating in the track. While MS MARCO V1 collection was used in the TREC 2019~\citep{Craswell:2019:Overview} and TREC 2020 DL~\citep{Craswell:2020:Overview} Tracks, the MS MARCO V2 collection was used in TREC 2021~\citep{Craswell:2021:Overview} and TREC 2022 \citep{Craswell:2022:Overview}. 
For both MS MARCO V1 and V2, we combine both available query sets to increase the pool of queries to sample from, resulting in 97 queries and 20,641 relevant passages across the combined query set for MS MARCO V1 (an average of 213 relevant passages per query) and 129 queries and 20,963 relevant passages for MS MARCO V2 (an average of 165 relevant passages per query)\footnote{We exclude the judgments for two development queries, for each of which there exist more than 40,000 relevant passages.}. 

For testing the performance of our reranking models fine-tuned on both deep and shallow datasets. We use the MS MARCO V1 Dev2 dataset\footnote{\url{https://ir-datasets.com/msmarco-passage.html\#msmarco-passage/dev/2}} and MS MARCO V2 Dev1 dataset\footnote{\url{https://ir-datasets.com/msmarco-document-v2.html\#msmarco-document-v2/dev1}}, which contain the same set of passages as the respective Train sets. The distribution of the relevant passages in these Test sets is the same as in the Train sets, with an average of one relevant passage per query. The Test sets can thus be also considered to be shallow.  

\subsection{LongEval}
The LongEval-Retrieval~\citep{Galuscakova:2023:SIGIR} collection focuses on evaluating the temporal persistence of IR systems by simulating an evolving Web environment. The LongEval collection consists of successive subsets, each representing the state of the information system at a particular time interval. Each subset includes a set of queries, documents, and soft relevance judgments derived from user click models based on data from the Qwant search engine\footnote{https://www.qwant.com/}. The statistics of the collections are displayed in Table \ref{tab:longeval:stats}.

We use the Train set to create \textbf{shallow} datasets. It consists of 1.5 million Web pages and 754 queries from the Train and Heldout sets, with 1,493 relevant documents (an average of two relevant documents per query). On the other hand, \textbf{deep} datasets are created using a pool of manually evaluated results from systems submitted to the LongEval task. These deep dataset features contain 50 Heldout queries with 5,793 associated relevant documents (an average of 116 relevant documents per query)\footnote{The Test collection is currently not publicly available, but will be provided in the GitHub.}. 

We use both available Test sets, which focus on different time shifts from the Train set. Both these Test sets have distributions similar to the Train and Heldout sets, with 1.5 million documents in the short-term (ST) dataset and 1 million documents in the long-term (LT) dataset. Each Test set contains around 900 queries and approximately 3,500 relevant documents (an average of four relevant documents per query). These two Test sets were used for measuring the performance of our reranking models, fine-tuned on both shallow and deep datasets.

\begin{table*}
\begin{tabular}{ll|cccc}
\hline
& &\multicolumn{4}{c}{\textbf{LongEval}} \\
\textbf{Type} & \textbf{Collection} & \textbf{Query} & \textbf{Qrels} & \textbf{PQrels} & \textbf{Docs} \\
\hline
\multirow{1}{*}{\textbf{Deep}} & Heldout Queries & 50 & 5,973 & 2,907 & 5,306 \\
\hline
\multirow{3}{*}{\textbf{Shallow}} & Train & 754 & 1493 & - &  1,570,734 \\
& Test ST & 882 & 12,217 & 3,370 & 1,593,376 \\
& Test LT & 923 & 13,467 & 3,835 & 1,081,334 \\

\hline
\end{tabular}
\caption {Statistics of the LongEval collection used in the experiments. Number of queries (Query), all relevance judgments (Qrels), positive relevance judgments (PQrels) and documents (Docs) are displayed. Test collections were used as Short-term (ST) and Long-term (LT).}
\label{tab:longeval:stats}
\end{table*}

\section{Experiments}
\label{sec:experiments}


To assess the reranker’s performance on both deep and shallow datasets, we construct several training sets with certain characteristics from the MS MARCO and LongEval collections. We then fine-tune BERT model from scratch on a sequence classification tasks using  these constructed sets and compare deep and shallow training data with matching number of training instances. In our experimental setup, both the retrieval and reranking stages are conducted at a depth of 10. Specifically, we use Pyserini\footnote{\url{https://github.com/castorini/pyserini}}~\citep{Lin:2021:SIGIR} to retrieve the top 10 documents for each query using BM25. These 10 documents are then passed to MonoBERT for reranking. All evaluation metrics (MAP, NDCG, MRR) are computed at rank 10, which aligns with our goal on assessing reranker performance on the most highly ranked results in these collections. 

\subsection{Reranking Model}
We use the \texttt{BertForSequenceClassification}\footnote{\url{https://huggingface.co/docs/transformers/en/model_doc/bert}} class from Hugging Face's library, as an implementation of the \texttt{bert-base-uncased} model. The models are fine-tuned on a Tesla V100 GPU and Tesla P100 GPU. Each document is tokenized using the BERT tokenizer and truncated or padded to 512 tokens. We set the batch size to $8$, the learning rate to $2e-5$, and train for 10 epochs. Training data are split on training and validation sets and early stopping, based on the F1 score and applied on the validation set, is used to prevent overfitting. Negative samples are for each query drawn from BM25 search results, with the top 10 results excluded to avoid potential false negatives.

\subsection{MS MARCO V1}
For both MS MARCO collections, we create corresponding deep and shallow subsets, each containing the same number of training instances. Specifically, in MS MARCO V1 we use 5,000 instances for training, with 50 \textbf{deep} queries and 2,500 \textbf{shallow} queries, along with their corresponding relevance judgments (i.e. 100 judgments for the deep queries and 2 judgments for the shallow queries). In both the deep and shallow sets, we maintain a 1:1 ratio of positive to negative samples. For example, in the shallow 2,500/2 set, each query has exactly one positive and one negative judgment.
We also apply the same setup with a slightly smaller number of training instances (4,200). In this case, we use 70 \textbf{deep} queries with 60 relevance judgments and corresponding 2,100 \textbf{shallow} queries with 2 relevance judgments. The results of these experiments are displayed in Table~\ref{tab:msmarco}. Alongside the BM25 baseline, we include results from a zero-shot approach using an pre-trained BERT model for reranking. 

The zero-shot approach underperforms the BM25 baseline, highlighting the importance of fine-tuning the BERT model for reranking.
The models fine-tuned on the shallow datasets significantly outperform those fine-tuned on the corresponding deep datasets, as well as the BM25 baseline. The deep datasets even perform significantly worse than the BM25 baseline, which may indicate that such a small number of training queries in the deep dataset can lead to overfitting. 

\begin{table*}[ht]
\centering
\begin{tabular}{lccclll}
\hline
\textbf{Model} & \textbf{Type} &  \textbf{Inst.} & \textbf{Q/Q Ratio} & \textbf{MAP} & \textbf{NDCG} & \textbf{MRR} \\ \hline
BM25 & {-} & {0} & {-} & 0.1793  & 0.2269  & 0.1852 \\
BM25+BERT & - & {0} & - & 0.1534*  & 0.2068*  & 0.1582* \\ 
BM25+BERT & Deep & 4,200 & 70/60 & 0.1136* & 0.1756* & 0.1183* \\
BM25+BERT & Deep & 5,000 & 50/100 & 0.1145* & 0.1748* & 0.1173* \\
BM25+BERT & Shallow & 4,200 & 2,100/2 & 0.2128* & 0.2578* & 0.2255* \\
BM25+BERT & Shallow & 5,000 & 2,500/2 & \textbf{0.2201*} & \textbf{0.2594*} & \textbf{0.2277*}\\
\hline
\end{tabular}
\caption{Scores for models fine-tuned and evaluated on the MS MARCO v1 collection, depending on the number of training instances (Inst.) and their Query/Qrels Ratio (Q/Q Ratio). The highest-performing models are highlighted in bold.
*: Indicates that the model's performance is significantly different from the BM25 baseline, as determined by a t-test with a significance threshold 0.05.}
\label{tab:msmarco}
\end{table*}

\subsubsection{Shallow Datasets}
Since the shallow datasets clearly outperform the deep ones, we further investigate how many shallow training instances are needed to surpass the BM25 baseline. To this end, we progressively reduce the number of training instances in the shallow datasets. Table~\ref{tab:msmarco-shallow} shows the relationship between model performance and the number of training instances in the shallow dataset, ranging from 500 to 5,000. The zero-shot BM25 baseline can be outperformed using approximately 1,000 training instances, which is slightly lower than the estimate reported by~ \citet{Nogueira:2020:ACL}. 

\begin{table*}[ht]
\centering
\begin{tabular}{lccclll}
\hline
\textbf{Model} & \textbf{Type} &  \textbf{Inst.} & \textbf{Q/Q Ratio} & \textbf{MAP} & \textbf{NDCG} & \textbf{MRR} \\ \hline
BM25 & {-} & {0} & {-} & 0.1793  & 0.2269  & 0.1852 \\
BM25+BERT & - & {0} & - & 0.1534*  & 0.2068*  & 0.1582* \\ 
BM25+BERT & Shallow & 500 & 250/2 & 0.1792 & 0.2273 & 0.1855\\
BM25+BERT & Shallow & 1,000 & 500/2 & 0.1993* & 0.2432* & 0.2061*\\
BM25+BERT & Shallow & 2,000 & 1,000/2 & 0.2110* & 0.2521* & 0.2180*\\
BM25+BERT & Shallow & 3,000 & 1,500/2 & 0.2158* & 0.2560* & 0.2233*\\
BM25+BERT & Shallow & 4,000 & 2,000/2 & 0.2108* & 0.2518* & 0.2179*\\
BM25+BERT & Shallow & 5,000 & 2,500/2 & \textbf{0.2201*} & \textbf{0.2594*} & \textbf{0.2277*}\\
\hline
\end{tabular}
\caption{Performance of the models trained using shallow datasets, trained and evaluated on the MS MARCO v1 collection, dependent on the number of training instances (Inst.) and their Query/Qrels Ratio (Q/Q Ratio). The highest scores are highlighted in bold. *: Indicates that the model’s performance is significantly different from the BM25
baseline, as determined by a t-test with a significance threshold 0.05.}
\label{tab:msmarco-shallow}
\end{table*}

\subsubsection{Negative Judgments}

Although positive judgments in IR are typically acquired using some human intervention, negative judgments are often generated automatically~\cite{Xiong:2021:ICLR}. This allows the number of negative judgments to be scaled up easily. Accordingly, we increase the number of training instances by including additional negative passages retrieved using BM25. While adding negative passages significantly improves the performance of fine-tuned models on the deep datasets, it has only a limited effect on the shallow datasets, as shown in Table \ref{tab:msmarco-ratios}. For models fine-tuned on deep-based datasets, performance peaks at a positive-to-negative sample ratio of 1:8, then declines as the ratio increases further to 1:10 and 1:20. Rerankers fine-tuned on shallow-based datasets show similar performance across ratios from 1:1 to 1:8. As with the deep datasets, performance decreases for higher ratios.

\begin{table*}[ht]
\centering
\begin{tabular}{lccclll}
\hline
\textbf{Type} & \textbf{Inst.} & \textbf{Q/Q Ratio} & \textbf{P/N Ratio} & \textbf{MAP} & \textbf{NDCG} & \textbf{MRR} \\ \hline
Deep & 5,000 & 50/100 & 1:1 & 0.1145 & 0.1756 & 0.1183 \\
Deep & 7,500 & 50/150 & 1:2 & 0.1315 & 0.1894 & 0.1357 \\ 
Deep & 12,500 & 50/250 & 1:4 & 0.1363 & 0.1930 & 0.1405 \\ 
Deep & 17,500 & 50/350 & 1:6 & 0.1607 & 0.2125 & 0.1655 \\ 
Deep & 22,500 & 50/450 & 1:8 & \textbf{0.1704} & \textbf{0.2202} & \textbf{0.1760} \\ 
Deep & 27,500 & 50/550 & 1:10 & 0.1627 & 0.2140 & 0.1681 \\ 
Deep & 52,500 & 50/1050 & 1:20 & 0.1652 & 0.2162 & 0.1706 \\ 
\hline
Shallow & 5,000 & 2,500/2 & 1:1 & 0.2201 & 0.2594 & 0.2277 \\ 
Shallow & 7,500 & 1,250/5 & 1:2 & 0.2068 & 0.2490 & 0.2139 \\ 
Shallow & 12,500 & 2,500/5 & 1:4 & \textbf{0.2233} & \textbf{0.2610} & \textbf{0.2298} \\ 
Shallow & 17,500 & 2,500/7 & 1:6 & 0.2203 & 0.2594 & 0.2274 \\
Shallow & 22,500 & 2,500/9 & 1:8 & 0.2219 & 0.2606 & 0.2290 \\
Shallow & 27,500 & 2,500/11 & 1:10 & 0.1474 & 0.2015 & 0.1521 \\
Shallow & 52,500 & 2,500/21 & 1:20 & 0.1612 & 0.2130 & 0.1668 \\
\hline
\end{tabular}
\caption{Performance for the models fine-tuned and evaluated on the MS MARCO v1 collection, dependent on the positive-to-negative sample ratios (P/N Ratio), number of training instances (Inst.) and their Query/Qrels Ratio (Q/Q Ratio). The highest scores for deep and shallow sets are highlighted in bold.}
\label{tab:msmarco-ratios}
\end{table*}

\subsection{MS MARCO V2}
Compared with MS MARCO V1, the available deep collections in MS MARCO V2 provide more queries and relevant passages, facilitating the construction of deep training sets.

We create the following \textbf{deep} training sets:
\begin{itemize}
    \item The first set consists of 50 queries, each with 100 judgments (deep 50/100).
    \item The second set has 70 queries, each with 60 judgments (deep 50/200).
    \item In the third set, we increase the number of queries to 100 and include 100 judgments (deep 100/100).
    \item In the fourth set, we again use 100 queries and increase the number of judgments to 140 (deep 100/140).
\end{itemize}
and corresponding \textbf{shallow} training sets:
\begin{itemize}
    \item The first set includes 2,500 queries, each with 2 judgments (shallow 2,500/2).
    \item The second set maintains 2,500 queries and we increase the number of judgments to 4 (shallow 2,500/4).
    \item In the the third set, we comprise 5,000 queries, each with 2 judgments (shallow 5,000/2).
    \item In the fourth set, we use 3,500 queries with 4 judgments (shallow 3,500/4).
\end{itemize}

Table \ref{tab:msmarco2} presents the results of experiments using a 1:1 ratio of positive to negative samples. The results are consistent with those observed in MS MARCO V1.
Experiments using deep judgments perform significantly worse than the BM25 baseline, whereas those using shallow judgments, covering a broader range of queries, outperform it.
The results also reveal a trend that the performance improves with the size of the training set. Interestingly, this trend is particularly noticeable with an increasing number of queries in the shallow sets (e.g., shallow 2,500/2 vs. shallow 5,000/2), but it also holds when increasing the number of positive samples per query (e.g., shallow 2,500/2 vs. shallow 2,500/4).

\begin{table*}[ht]
\begin{tabular}{lccclll}
\hline
\textbf{Model} & \textbf{Type} &  \textbf{Inst.} & \textbf{Q/Q Ratio} & \textbf{MAP} & \textbf{NDCG} & \textbf{MRR} \\ \hline
BM25 & {-} & {0} & {-} & 0.0642  & 0.0839  & 0.0651 \\
BM25+BERT & - & {0} & - & 0.0323  & 0.0582  & 0.0328 \\ 
BM25+BERT & Deep & 5,000 & 50/100 & 0.0462* & 0.0697* & 0.0468* \\
BM25+BERT & Deep & 10,000 & 50/200 & 0.0420* & 0.0663* & 0.0427* \\
BM25+BERT & Deep & 10,000 & 100/100 & 0.0443* & 0.0682* & 0.0449* \\
BM25+BERT & Deep & 14,000 & 100/140 & 0.0507*  & 0.0732*  & 0.0513* \\ 
BM25+BERT & Shallow & 5,000 & 2,500/2 & 0.0664 & 0.0858 & 0.0671 \\
BM25+BERT & Shallow & 10,000 & 5,000/2 & 0.0699* & 0.0885* & 0.0707* \\                              
BM25+BERT & Shallow  & 10,000 & 2,500/4 & 0.0674 & 0.0866 & 0.0682 \\ 
BM25+BERT & Shallow & 14,000 & 3,500/4 & \textbf{0.0702*} & \textbf{0.0889*} & \textbf{0.0770*} \\ \hline
\end{tabular}
\caption{Performance of the models fine-tuned and evaluated on the MS MARCO V2 collection, depending on the number of training instances (Inst.) and their Query/Qrels Ratio (Q/Q Ratio). The highest scores are highlighted in bold.
*: Indicates that the model's performance is significantly different from the BM25 baseline, as determined by a t-test with a significance threshold 0.05.}
\label{tab:msmarco2}
\end{table*}

\subsubsection{Negative Judgments}
Similar to the MS MARCO V1 collection, we also experiment with increasing the ratio of positive to negative samples. Specifically, we raise the positive-to-negative sample ratio to 1:4 for the deep 50/100 and deep 50/200 datasets, as well as their corresponding shallow 2,500/2 and 5,000/2 datasets. As shown in Table~\ref{tab:msmarco2-ratios}, training sets with a 1:4 ratio yield improved performance compared to those with a 1:1 ratio.

Including more negative samples helps the model learn to identify non-relevant documents, enhancing its ability to distinguish between relevant and non-relevant content. These results suggest that increasing the number of negative samples per query can partially mitigate the limitations caused by a scarcity of positive judgments. While this effect was observed only in the deep datasets of the MS MARCO V1 collection, it holds for both shallow and deep datasets in the MS MARCO V2 collection.


\begin{table*}[ht]
\centering
\begin{tabular}{lccclll}
\hline
\textbf{Type} & \textbf{Inst.} & \textbf{Q/Q Ratio} & \textbf{P/N Ratio} & \textbf{MAP} & \textbf{NDCG} & \textbf{MRR} \\ \hline
Deep & 5,000 & 50/100 & 1:1 & 0.0462 & 0.0697 & 0.0468 \\
Deep & 12,500 & 50/250 & 1:4 & 0.0524 & 0.0747 & 0.0531 \\ 
Deep & 10,000 & 50/200 & 1:1 & 0.0420 & 0.0663 & 0.0427 \\ 
Deep & 25,000 & 50/500 & 1:4 & 0.0540 & 0.0760 & 0.0547 \\ 
Shallow & 5,000 & 2,500/2 & 1:1 & 0.0664 & 0.0858 & 0.0671 \\ 
Shallow & 12,500 & 2,500/5 & 1:4 & 0.0729 & 0.0908 & 0.0737 \\ 
Shallow & 10,000 & 5,000/2 & 1:1 & 0.0699 & 0.0885 & 0.0707 \\ 
Shallow & 25,000 & 5,000/5 & 1:4 & \textbf{0.0803} &   \textbf{0.0916} & \textbf{0.0813} \\ \hline
\end{tabular}
\caption{Performance of the models fine-tuned and evaluated on the MS MARCO~V2 collection with 1:1 and 1:4 positive-to-negative sample ratios (P/N Ratio), depending depending on the number of training instances (Inst.) and their Query/Qrels Ratio (Q/Q Ratio). The highest scores are indicated in bold.}
\label{tab:msmarco2-ratios}
\end{table*}

\begin{table*}[ht]
\centering
\begin{tabular}{lcccclll}
\hline
\textbf{Test} &
\textbf{Model} & \textbf{Type} &
\textbf{Inst.} &
\textbf{Q/Q Ratio} &
\textbf{MAP} & \textbf{NDCG} & \textbf{MRR} \\ \hline
\multirow{5}{*}{\textbf{ST}} & BM25 & - & - & - & \textbf{0.1189} & 0.1746 & 0.2430\\
& BM25+BERT & - & - & - & 0.1183 & \textbf{0.1749} & \textbf{0.2474}\\
& BM25+BERT & Deep & 2,250 & 45/50 & 0.1034* & 0.1607* & 0.2068*\\
& BM25+BERT & Deep & 3,100 & 31/100 & 0.1155* & 0.1709* & 0.2324*\\
& BM25+BERT & Shallow & 1,508 & 754/2 & 0.1064* & 0.1637* & 0.2199*\\
\hline
\multirow{5}{*}{\textbf{LT}} & BM25 & - & - & - & \textbf{0.1150} & \textbf{0.1736} & \textbf{0.2505}\\
& BM25+BERT & - & - & - & 0.0702* & 0.1297* & 0.1337*\\
& BM25+BERT & Deep & 2,250 & 45/50 & 0.1027* & 0.1618* & 0.2208*\\
& BM25+BERT & Deep & 3,100 & 31/100 & 0.1103 & 0.1681 & 0.2357\\
& BM25+BERT & Shallow & 1,508 & 754/2 & 0.1070* & 0.1654* & 0.2339\\
\hline
\end{tabular}
\caption{Performance of the models fine-tuned and evaluated on the LongEval collection, depending on the number of training instances (Inst.) and their Query/Qrels Ratio (Q/Q Ratio). Evaluation is done on the Short-term (ST) and Long-term (LT) LongEval test sets. The highest scores on the ST/LT test set are indicated in bold.
*: Indicates that the model's performance is significantly different from the BM25 baseline based on a t-test with a significance threshold $0.05$.}
\label{tab:longeval}
\end{table*}


\subsection{LongEval}
To directly compare deep and shallow datasets, we require collections that contain both types of setups. Another such collection is LongEval. 
For LongEval, we follow the same setup as used for MS MARCO, although the number of both deep and shallow queries is limited. We use all 754 queries available in the Train set as a shallow set. For the deep set, we have 45 queries, each with 25 manually collected relevant documents, and 31 queries, each with 50 manually collected relevant documents. Thus, the number of training instances in the largest available shallow dataset is smaller than the number of instances in the available deep datasets.

Table~\ref{tab:longeval} summarizes the performance on the ST and LT test sets. In all experiments, we apply a 1:1 positive-to-negative sampling ratio. 
In none of the configurations is the number of training instances sufficient to outperform the BM25 baseline. Nevertheless, these setups allow us to further compare the effects of training data type and volume.

The model trained on the shallow dataset with 1,508 training instances performs better than the model trained on the deep dataset with 2,250 training instances, and worse than the one trained on the deep dataset with 3,100 training instances—across both test sets and all evaluation measures. Though the absolute score differences are small, they are statistically significant. These experiments confirm that deep datasets require several times more training instances than shallow datasets to achieve comparable performance.

\section{Discussion}
\label{sec:discussion}
Not surprisingly, as we increase the number of training samples in both deep and shallow datasets, reranker models trained on larger datasets exhibit improved performance. However, our main observations regarding shallow versus deep datasets, based on the presented results, are as follows:

\begin{itemize}
    \item A significantly smaller number of positive samples is needed in shallow datasets compared to deep datasets to achieve similar performance when training a reranking model.   
    \item Negative samples help mitigate the lack of training data, particularly in deep datasets. Adding negative samples to the training sets boosts performance, up to a certain positive-to-negative sample ratio. 
\end{itemize}

One of the main limitations of this study is the lack of suitable train collections and training data. As it is rare to have both deep and shallow datasets available for a single collection, our conclusions cannot be easily repeated on additional collections. Moreover, while performance on deep collections clearly improves with more queries, such collections are often not large enough to outperform the baseline. Although our results on the MS MARCO V1 and V2 collections reflect similar trends, MS MARCO V2 may contain many unjudged relevant documents, making it unreliable for evaluation ~\cite{Frobe:2022:CLEF,Voorhees:2022:SIGIR}. 

The number of queries in deep collections is typically set to 50, as this has proven sufficient for measuring significant differences between IR systems~\cite{Manning:2008:book}. Nevertheless, we hypothesize that increasing this number could make such collections suitable not only for evaluation but also for effective training of IR systems, especially when coupled with an appropriate negative sampling strategy. In our experiments, we used a BERT model as a prototype reranker, but more advanced models might require even fewer training instances~\cite{Nogueira:2020:ACL}. If combined with findings by~\citet{Lu:2017:SIGIR}, we might cautiously suggest that a slight `shallowing' of traditional deep evaluation collections could help create datasets useful for both evaluation and training of neural IR models while staying within the same budget. 



\section{Conclusion}
\label{sec:conclusion}


Shallow training sets consistently outperform deep training sets in all our experiments. Shallow dataset require much less training instances to achieve similar performance as the deep datasets. Thus, we confirm that the broader topic coverage in shallow training sets enables models to generalize better and perform more effectively in reranking tasks. Increasing the size of shallow training sets further greatly enhances reranking performance. The issue of lack of training data can be partially mitigated by increasing the number of negative training samples, what might be helpful especially for deep datasets. Albeit it is important to monitor the ratio of positive to negative samples, as a highly imbalanced proportion might add noise to the training data. 

\bibliographystyle{ACM-Reference-Format}
\balance
\bibliography{ictir25paper}

\onecolumn
\section*{Appendix}
\label{sec:appendix}

\begin{figure*}[h]
  \centering
  \includegraphics[width=0.8\textwidth]{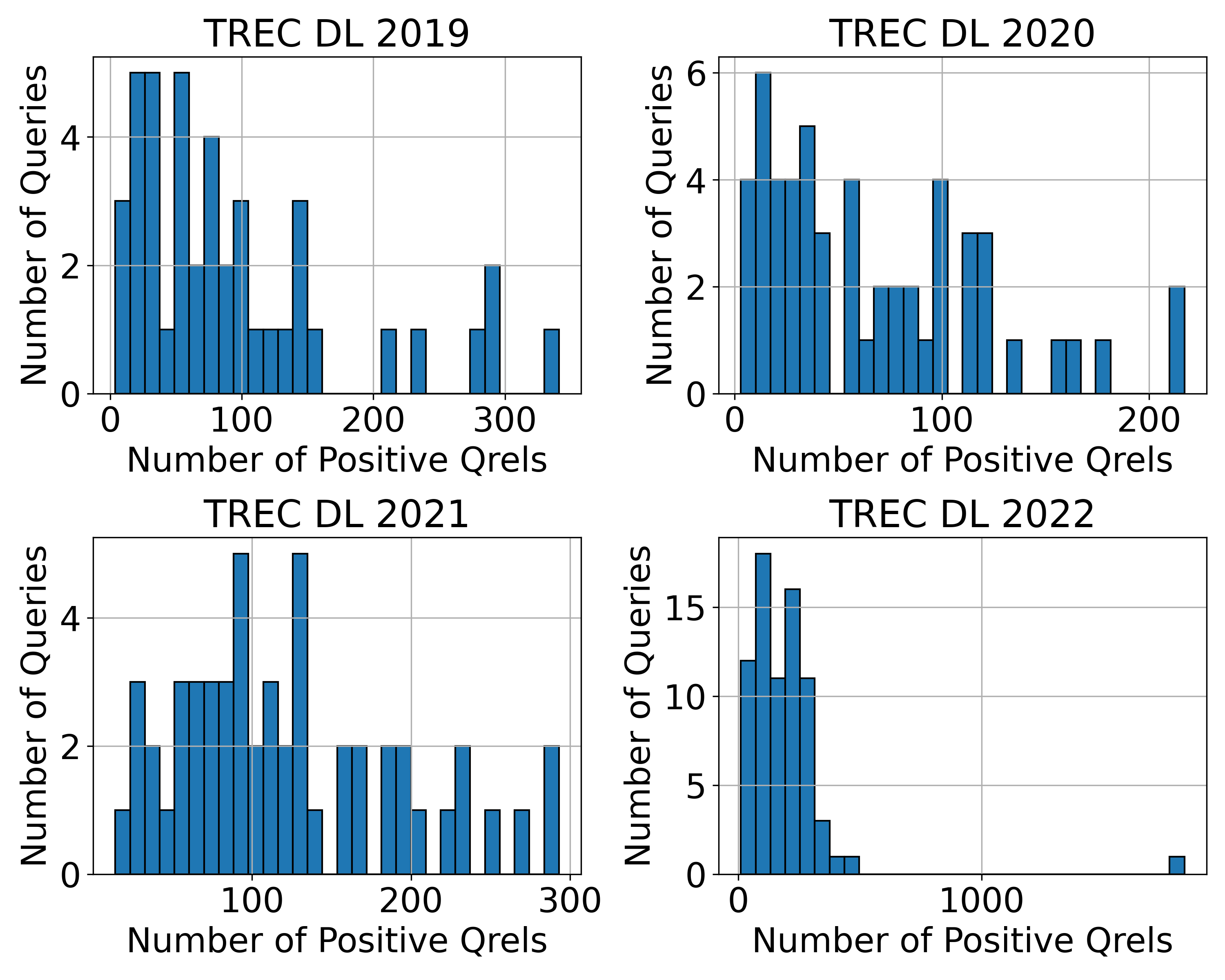}
  \caption{\textbf{Distribution of the number of positive relevance passages (Qrels) per query in the TREC DL datasets from 2019-2020 (MS MARCO V1) and 2011-2022 (MS MARCO V2). Each histogram illustrates the variation in the number of relevant passages across queries. We consider these datasets to be deep datasets.}}
  \label{fig:trec-dl-histograms}
\end{figure*}

\begin{figure*}[h]
  \centering
  \includegraphics[width=0.8\textwidth]{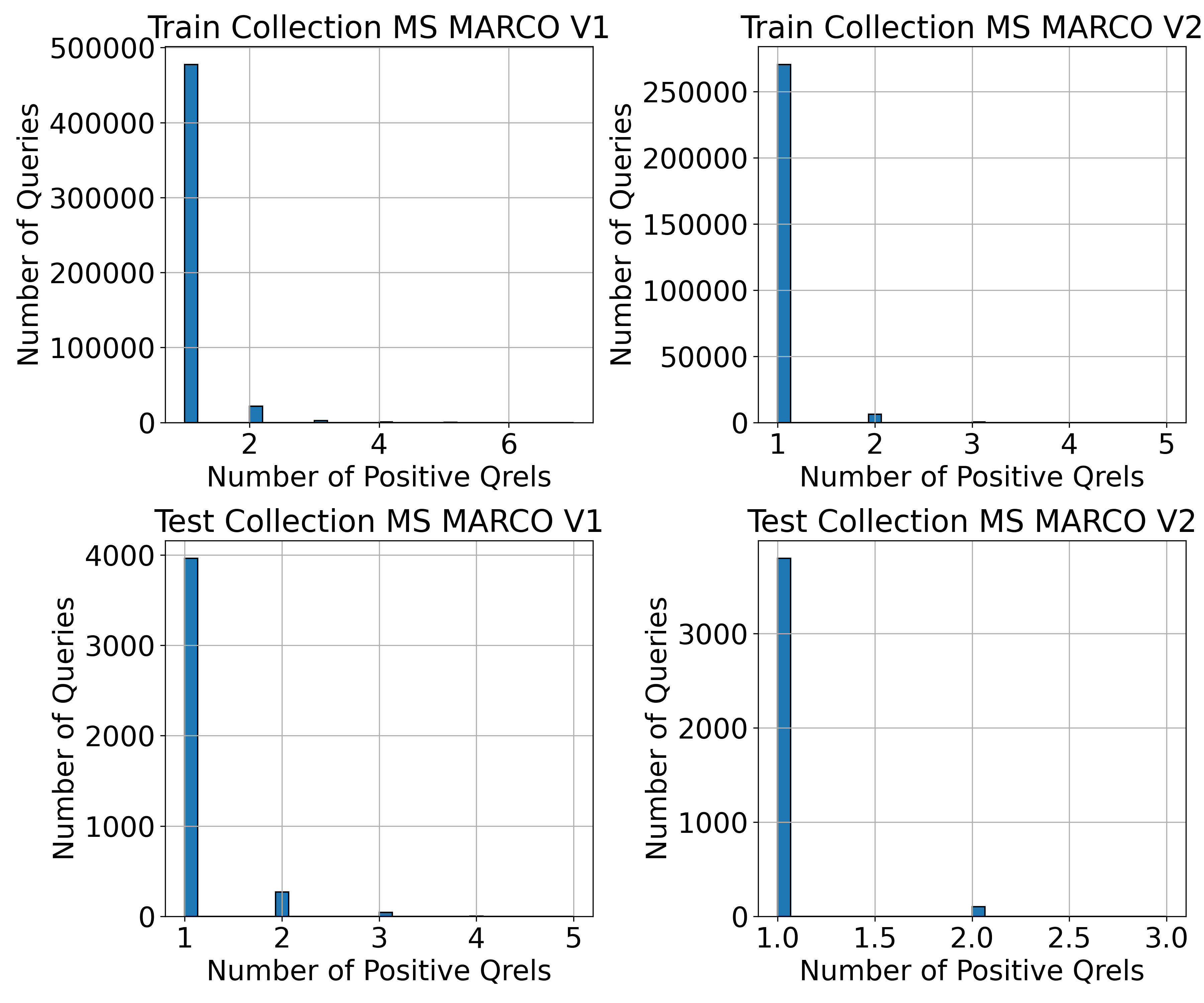}
  \caption{\textbf{Distribution of the number of positive Qrels per query in the MS MARCO V1 and V2 collections, for both training and test sets. We consider these datasets to be shallow datasets.}}
  \label{fig:train-test-MS-MARCO}
\end{figure*}

\begin{figure*}[h]
  \centering
  \includegraphics[width=0.8\textwidth]{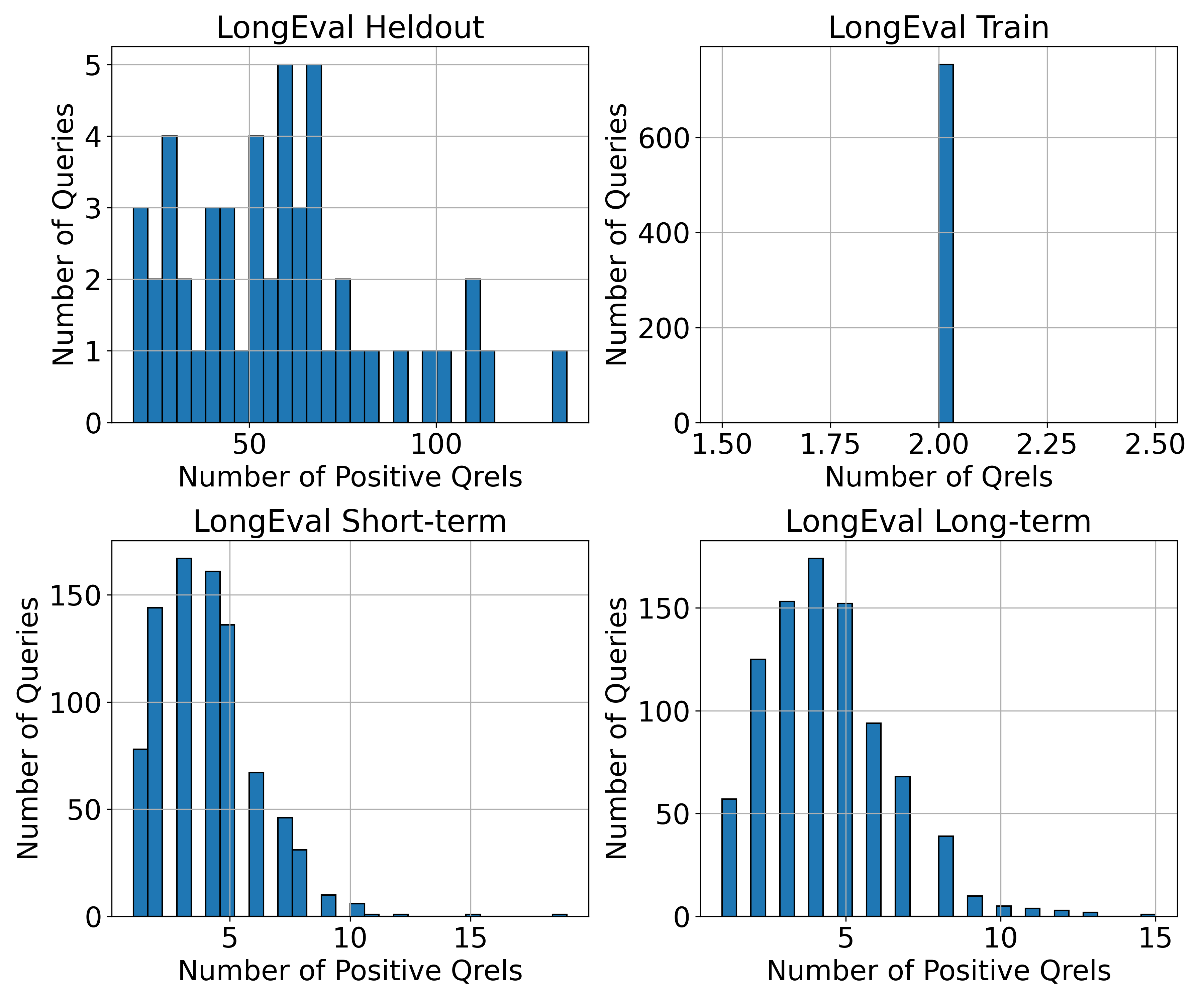}
  \caption{\textbf{Distribution of the number of positive relevance passages (Qrels) per query in the LongEval datasets. Each histogram illustrates the variation in the number of positive relevant passages, except for the LongEval train, which depicts only relevant passages across queries. We consider the heldout dataset to be the deep dataset and the train and test sets to be shallow datasets.}}
  \label{fig:longeval-histogram}
\end{figure*}

\twocolumn  

\end{document}